# Unusual Behaviour of (Np,Pu)$B_2C$

Tomasz Klimczuk[1,2], Pascal Boulet[3], Jean-Christophe Griveau[3], Eric Colineau[3], Ernst Bauer[4], Matthias Falmbigl[5], Peter Rogl[5] and Franck Wastin[3]

[1]*Faculty of Applied Physics and Mathematics, Gdansk University of Technology, Narutowicza 11/12, 80-233 Gdansk, Poland*

[2] *Institute of Physics, Pomeranian University, Arciszewskiego, 76-200 Slupsk, Poland*

[3]*European Commission, Joint Research Centre, Institute for Transuranium Elements; Postfach 2340, 76125 Karlsruhe, Germany*

[4]*Institute of Solid State Physics, Vienna University of Technology, A-1040 Wien, Austria*

[5]*Institute for Physical Chemistry, University Vienna, Währingerstrasse 42, A-1090 Wien, Austria*

## Abstract

Two transuranium metal boron carbides, $NpB_2C$ and $PuB_2C$ have been synthesized by argon arc melting. The crystal structures of the {Np,Pu}$B_2C$ compounds were determined from single crystal X-ray data to be isotypic with the $ThB_2C$-type (space group $R\bar{3}m$, a = 0.6532(2) nm; c = 1.0769(3) nm for $NpB_2C$ and a = 0.6509(2) nm; c = 1.0818(3) nm for $PuB_2C$; Z=9). Physical properties have been derived from polycrystalline bulk material in the temperature range from 2 K to 300 K and in magnetic fields up to 9 T. Magnetic susceptibility and heat capacity data indicate the occurrence of antiferromagnetic ordering for $NpB_2C$ with a Neel temperature $T_N$ = 68 K. $PuB_2C$ is a Pauli paramagnet most likely due to a strong hybridisation of s(p,d) electrons with the Pu-*5f* states. A pseudo-gap, as concluded from the Sommerfeld value and the electronic transport, is thought to be a consequence of the hybridisation. The magnetic behaviour of {Np,Pu}$B_2C$ is consistent with the criterion of Hill.

**E-mail address:** peter.franz.rogl@univie.ac.at, tomasz.klimczuk@pg.gda.pl

**Corresponding author:** FAX: +48 58 347 2821

# 1. Introduction

The unique physical and chemical behaviour of actinoid (An) materials is essentially due to the unusual electronic structure combining features of the *d*-metal series with those of *f*-transition elements. Particularly in light actinides (up to Pu) the *5f*-wave functions play the role that *d*-wave functions have in the *d*-transition metal series, forming broad and hybridised *5f* - bands [1, 2, 3].

Earlier investigations of phase relations (see ref. 4 and review given in ref. 5) in the ternary combinations {Th,U,Np,Pu}-B-C prompted the existence of two sets of ternary compounds (i) {Th,U,Np,Pu}BC and (ii) {Th,U,Np,Pu}$B_2C$, the latter crystallizing with a unique rhombohedral structure type, $ThB_2C$ (space group $R\bar{3}m$), deriving from simple hexagonal $AlB_2$-type. ß-$UB_2C$, for which the $ThB_2C$-type is the high temperature modification, was found to undergo on cooling a phase transformation at 1675°C to a structurally closely related orthorhombic low temperature polymorph α-$UB_2C$ (space group *Pmma*)[6]. Preliminary susceptibility measurements revealed weak ferromagnetic behaviour for ß-$UB_2C$, whilst α-$UB_2C$ was found to be temperature independent paramagnetic [6]. From a more recent and detailed study of polycrystalline ß-$UB_2C$ magnetic susceptibility, magnetisation, magnetic structure (from neutron data), electric resistivity, magnetoresistance, specific heat, thermoelectric power and muon spectroscopy have been reported classifying ß-$UB_2C$ as a weak ferromagnet ($T_C$ ~ 75 K) with an enhanced electronic specific heat coefficient pointing towards *5f* electrons existing in two different manifestations, itinerant and localized [7,8].

As no investigations have hitherto been carried out to elucidate the physical properties of the Np- and Pu- homologues, a detailed inspection of the physical behaviour of {Np,Pu}$B_2C$ compounds became the subject of the present work.

## 2. Experimental

Polycrystalline ingots were obtained by arc melting stoichiometric amounts of the constituent elements under an atmosphere of high purity argon on a water-cooled copper hearth, using a Zr-getter alloy. Starting materials were used in the form of pieces as supplied by Merck AG (boron and graphite), and electro-refined 3N-neptunium and 3N6-plutonium metal supplied by Los Alamos National Laboratory. In order to ensure homogeneity, the arc-melted buttons were turned over and remelted 3 times, with weight losses below 0.5%. The samples were checked by X-ray powder diffraction data (Cu $K_\alpha$ radiation) collected on a Bragg-Brentano Siemens D500 and D8-diffractometer using a step size in 2Θ of 0.02 degrees. The diffraction patterns were analyzed by a Rietveld-type profile refinement method using the Fullprof program [9]. Single crystal X-ray diffraction data for {Np,Pu}$B_2$C were collected on an Enraf-Nonius CAD-4 four circle diffractometer employing monochromated Mo$K_\alpha$ radiation. Data were processed employing the Molen package [10]. Bulk magnetic measurements were performed on encapsulated chunks using a Quantum Design MPMS7-SQUID in fields up to 7 Tesla and in the temperature range from 2 to 300 K. The electrical resistivity was determined using a standard four-probe dc technique, with four 0.05-mm-diameter platinum wires glued to the sample using silver epoxy (Epotek H20E). Specific heat data were measured by using the relaxation method (Quantum Design PPMS) in the temperature range from 2 to 300 K (up to 9 T). Since our compounds contain highly radioactive Np or Pu, the small samples with the mass below 5 mg, were covered/wrapped in the heat conducting STYCAST 2850 FT resin. The mass of the resin was known and the raw experimental data have been corrected for this additional heat capacity contribution.

## 3. Results and Discussion

### 3.1. Structural Chemistry

X-ray single crystal counter data, obtained from small crystal fragments, which were broken from the arc melted specimens of {Np,Pu}$B_2$C, unambiguously revealed crystal symmetry, systematic extinctions and cell size as characteristic for the Th$B_2$C-type (see Table 1). No supercell reflections were observed. Intensity data refinement (with anisotropic atom displacement factors for the metal atoms and isotropic thermal parameters for the non-metal atoms) showed full occupation of all atom sites in agreement with the formula An$B_2$C. The low residual values obtained confirmed isotypism with the structure type of Th$B_2$C [11]. Final residual electron densities were below 4 and 7 e/Å³ for $NpB_2C$ and $PuB_2C$, respectively. For interatomic distances see Table 1; listings of $F_o$ and $F_c$ values can be obtained on request.

The compounds of {Np,Pu}$B_2C$ are further representatives of a typical non-metal layered metal boron carbide structure, where non-metals form planar nets, $6B(6B3C)^2$, sandwiching puckered hexagonal metal sheets (see Figure 1). Comparison with the covalent atom radii ($R_B$=0.088, $R_C$=0.077 nm) yields typical covalent single B-B bonding ($d_{B-B}$=0.180 nm), but a double bond character for the short B-C bond distances, $d_{B-C}$=0.150. There are no direct carbon-carbon interactions in the structure. Carbon atoms are framed by four metal atoms in planar co-ordination, with two short double bonds to boron atoms forming a total co-ordination figure of a tetragonal bi-pyramid $[An_4B_2]C$. As typical for most metal borides [12], boron atoms centre a triangular prism of actinoid metal atoms $[An_6]B$ with each boron atom forming along the waist of the prism three more (covalent) bonds to two B atoms and one C-atom. Whereas metal-nonmetal bonds essentially correspond to the sum of radii, metal-metal interaction (seven bonds for An1-atoms and eight bonds for An2-atoms) reveals only few bonds at a distance slightly shorter that the sum of metal (12-fold coordination) radii. These short bonds (two An1-An2 and one An2-An1 distances i.e. $d_{Np1-Np2}$=0.3363 nm, $d_{Pu1-Pu2}$=0.3380 nm respectively) indicate increased *f-d* hybridisation.

Characterization of bulk material by Rietveld refinements revealed practically single-phase conditions with only small amounts of secondary phases, i.e. $PuB_4$ and traces of PuBC in $PuB_2C$ ($R_F$ = 0.062) and traces of NpBC in $NpB_2C$ ($R_F$ = 0.065), respectively. Although the amount of these phases is less than a few mass percent, they may influence magnetic and transport properties of the bulk material (see below). Both compounds {Np,Pu}BC are isotypic with the UBC-type [4, 5].

## 3.2. Physical Properties

Figure 2(a) presents the temperature dependent magnetic susceptibility, $\chi(T)=M/H$, for $NpB_2C$, which reveals typical Curie-Weiss-like behaviour and is characterized by an antiferromagnetic transition at $T_N \sim 75$ K without a significant variation of $T_N$ with field (not shown here). In order to estimate the precise value of the Néel temperature, $d(\chi T)/dT$ (solid blue line) is plotted on the same temperature scale [13]. The Néel temperature is defined by the maximum of $d(\chi T)/dT$ and for $NpB_2C$ $T_N$ = 68 K is derived. An increase of $\chi$ below 50 K is likely caused by the presence of a ferromagnetic impurity phase. This phase is neither $NpB_2$ ($T_C \sim 99.5$ K [14]) nor $NpB_4$ ($T_N$ ~ 52.5 K [14]) but very likely corresponds to ferromagnetic NpBC with $T_C$ ~ 60 K [15], which is seen in small amounts in the X-ray patterns of $NpB_2C$ (see section Structural Chemistry).

The high temperature part of the magnetic susceptibility ($\chi$ = M/H) can be fitted to the modified Curie-Weiss law, $\chi = \chi_0 + C/(T-\Theta_p)$, where $\chi_0$ is a temperature independent term, $C$ is the Curie constant and $\Theta_p$ is the paramagnetic Curie temperature. The temperature dependence of the inverse susceptibility for $NpB_2C$ together with a fit to the Curie-Weiss law (solid line), are shown in the inset of Figure 2. From this data, the effective moment per Np can be obtained from $\mu_{eff} = (8C)^{1/2}$ = 1.65 $\mu_B$/Np. This value is much smaller compared to the expected free ion value of $Np^{+3}$ (2.87 $\mu_B$) and the value observed for NpFeAsO [16] (2.78 $\mu_B$). However, a reduced effective paramagnetic moment is often observed, e.g. $\mu_{eff}$ = 1.5 $\mu_B$/Np for $NpCoGa_5$ [17], and can be caused by strong hybridization of the 5*f*- electrons with the conduction band. The positive value of the Curie-Weiss temperature, $\Theta_p$ = 27 K, might indicate the presence of ferromagnetic interactions in the compound. However, this effect is more likely caused by the presence of a small amount of the ferromagnetic impurity phase, e.g. NpBC, in the sample. The temperature independent susceptibility for $NpB_2C$ ($\chi_0 = 531\times10^{-6}$ emu/mol) is very similar to that observed for ß-$UB_2C$ ($\chi_0 = 600\times10^{-6}$ emu/mol) [7].

A recent $^{237}$Np-Mößbauer spectroscopy study on $NpB_2C$ [18] confirmed the magnetic ordering of both Np-sublattices below about 80 K with average ordered moments of $\langle\mu_{ord}\rangle$(Np,3a) = 0.30 $\mu_B$ and $\langle\mu_{ord}\rangle$(Np,6c) = 1.05 $\mu_B$, respectively at 4 K. The small magnetic moments and isomer shifts (as compared to the $Np^{3+}$ ground state, J=4) for Np in site 3a (IS = 9.8 mm/s) and in site 6c (IS = 0.1 mm/s) were interpreted [18] in favor of hybridization of the 5f Np-shell with the conduction band leading to a rather delocalized 5f magnetism in $NpB_2C$ similar to the predominantly itinerant ferromagnetism in isotypic ß-$UB_2C$ [7,8].

In contrast to $NpB_2C$, the magnetic susceptibility $\chi$(T) for $PuB_2C$ indicates a practically temperature independent paramagnetism (TIP). As can be seen from Figure 2(b), the relative change of $\chi$(T) from 300 K to 100 K is less than 10%, whereas for $NpB_2C$ in the same temperature range the relative change of $\chi$(T) is about 200%. Below 30 K the susceptibility rapidly increases and then saturates below 10 K. In order to shed light on this transition, magnetisation versus field, M(H), up to $\mu_0H$ = 7 T was measured at 2 K and 300 K (see inset of Figure 2(b)). M(H) yields a small spontaneous magnetisation of $2.5*10^{-3}$ $\mu_B$/Pu-atom and $2*10^{-3}$ $\mu_B$/Pu-atom at 2 K and 300 K, respectively. A linear dependence of M(H) in the applied magnetic field range from 2 to 7 Tesla is observed at 2 K and at 300 K. The magnetization at T=300 K, and $\mu_0H$ = 7 T, is only 0.006 $\mu_B$/Pu-atom. This magnetic behaviour is apparently not directly related to the intrinsic magnetic properties of $PuB_2C$, and may be attributed to parasitic ferromagnetic impurities brought about from Fe-impurities when crushing the boron starting material to fine powder in a steel mortar. A simple estimation on the basis of an assumed saturation moment of 1.5 $\mu_B$/Fe-atom shows that a small amount of about 500 ppm Fe per mol of Pu would already account for the observed magnetism. This amount is well below the X-ray detection limit. In addition the rise of $\chi(T)$ below 30 K may be due to a second ferromagnetic phase such as a ternary $PuFe_x(B,C)_y$. As Pu-borides and particularly $PuB_4$ were reported to be temperature independent paramagnets above 2 K (ref. 14), any magnetic influence of $PuB_4$ (small amounts in $PuB_2C$ – see section Structural Chemistry) can safely be ruled out. In conclusion, $PuB_2C$ is a temperature independent paramagnet with $\chi_0 < 450\times10^{-6}$ emu/mol.

The temperature dependent electrical resistivity, $\rho(T)$, at zero and externally applied magnetic fields is shown in Figs. 3(a,b). Both, the Np and the Pu based compounds exhibit a rather complicated $\rho(T)$ dependence and magnetic fields, obviously, do not change distinctly the absolute values of the resistivity as well as the overall shape.

$NpB_2C$ is characterised by a decreasing resistivity upon an increasing temperature in almost the entire temperature range inspected, i.e., $d\rho/dT < 0$. Since the Sommerfeld value of the specific heat is relatively large (see the following section), it cannot be concluded that the material is a semiconductor or next to a metal-to-insulator transition. As indicated from the magnetic susceptibility, a magnetic phase transition occurs around $T_N$ = 68 K. This can be read-off from $\rho(T)$ as well, from a distinct change of curvature in this temperature range. Below 68 K, $d\rho/dT$ is much larger than above the magnetic phase transition. This might indicate that the steeper increase below $T_N$ is a consequence of the onset of magnetic order at this temperature. In fact, various antiferromagnetic materials (see e.g., ref. 19) exhibit an unexpected increase of the electrical resistivity when crossing the transition temperature from above. This phenomenon is called super-zone boundary effect and is a result of the fact that magnetic ordering has a periodicity different from that of the lattice. As a consequence, new Brillouin zone boundaries are introduced, which come along with a gap in the electronic structure. The thus reduced number of charge carriers causes then an increase of the electrical resistivity. In general, however, at temperatures fairly below $T_N$ the resistivity starts to decrease because of the growing spatial and temporal order of the magnetic structure. The enhanced long-range periodicity is subsequently responsible for a distinct decrease of the interaction of conduction electrons with the magnetic moments of Np. As a result, a maximum is formed in $\rho(T)$ below the AFM phase transition temperature. In the present case, however, the electrical resistivity of $NpB_2C$ rises in a broad range below $T = T_N$ upon lowering the temperature.

The overall resistivity of $NpB_2C$ is of the order of several thousand μΩcm. This is unusually large and about 10 times larger than the respective values of $UB_2C$ [5]. A dramatically reduced charge carrier density or a substantial number of lattice defects can be made responsible for this observation. Weak localization effects of conduction electrons in intermetallic compounds are one of the rare additional mechanisms, able to explain a negative slope of dρ/dT in a broad temperature range. Localization is expected to be triggered in a metallic material by crystalline disorder such as site interchanges or the presence of vacancies in the crystal structure. Taking into account such a scenario, we have recently successfully explained similar temperature dependencies of ρ(T) for the Laves phases in the Mn-Cu-Si and Mn-Ni-Si systems [20]. In order to model ρ(T) in terms of weak localization, a set of dimensional dependent equations, worked out by Lee and Ramakrishnan [21], are considered. For the 3-dim case, this equation reads

$$\sigma = \sigma_0 + A^*T^{p/2} . \qquad (1)$$

Fitting the data of Fig. 3(a) in terms of Eqn. (1) for temperatures $T < T_N$, reveals p ~ 3. Here, p is an exponent used to parameterize the temperature dependence of the inelastic scattering length $L_i$ through $L_i \sim T^{-p/2}$. A is a constant representing some microscopic length scale of the problem, such as the inverse Fermi wave number. p = 2 would typically refer to predominant electron-phonon interactions in the system, while p = 3 possibly refers to electrons predominantly scattered on magnetic moments. Moreover, the applicability of Eqn. 1 even to the high temperature data indicates that there is no activation type behaviour in this system; thus a gap in the electronic density of states is not expected to be located near to the Fermi energy.

The electrical resistivity of $PuB_2C$ is about 4 to 5 times smaller than that of the Np based sample; ρ(T), however, behaves entirely uncommon for a metallic system. There are ranges with positive and negative dρ/dT dependences together with pronounced structures between 2 K and room temperature. Again, the Sommerfeld value $\gamma = 14$ mJ/mol$K^2$ refers to a metallic ground state of the system, but being 5 times smaller than in the case of magnetic $NpB_2C$. Within the actinide series, plutonium is located at the itinerant/localized boundary between the strongly hybridized *5f* states of U and localized *5f* states of Am. The rather small Sommerfeld value of $PuB_2C$, however, indicates that strong correlations among electrons are almost absent; thus it is not expected that features like the Kondo effect play a major role in this compound. Consequently it is not likely that the low temperature increase observed for ρ(T) is due to Kondo-type interaction. This conclusion is further supported by magnetoresistance data (see Fig 3(c)). While magnetic fields in typical Kondo systems suppress spin fluctuations and hence the electrical resistivity decreases, the data derived for $PuB_2C$ show an increase upon increasing magnetic fields. However, a standard, quadratic field dependence of the magnetoresistance, at least in the low field limit, is absent in this compound; rather, an almost linear temperature dependence of $\rho(H)$ is obvious for various temperatures. A least squares fit applied to the 15 K run (solid line, Fig. 3(c)) reveals an exponent n ~ 1.05. Various models are described in literature (compare e.g., ref. 22) that account for a linear dependence of $\rho(H)$. The residual resistivity of $PuB_2C$ is more than 100 times larger than typical simple intermetallic compounds and even more than 1000 times larger than simple metals like Cu or Ag. For such materials the classical magnetoresistance varies proportional to $H^2$ in the low field limit. The short mean free path of electrons in $PuB_2C$ that favours the classical regime, $\omega_c\tau << 1$ ($\omega_c$ and $\tau$ are the cyclotron frequency and the relaxation time of scattering, respectively), with a resulting quadratic field dependence, would even enlarge the field range, where a quadratic field dependence is observed. A non-classical origin of the distinct magnetoresistance behaviour in $PuB_2C$ is thus concluded.

In an attempt to get, at least, a rough idea about the uncommon $\rho(T)$ dependence, we have assumed that the Fermi energy of $PuB_2C$ is near to the edge of an electronic band; above this band edge there is narrow gap below further unoccupied states. In order to have relatively simple assumptions, the electronic bands are box-shaped like. The corresponding model has been described in ref. 23. A least squares fit according to such a model is shown as a solid line in Fig. 3(b).

The temperature dependence of the heat capacity ($C_p$) for $NpB_2C$ and $PuB_2C$ is shown in Figure 4(a) and Figure 4(b), respectively. The raw data were corrected for the additional $C_p$ signal of the stycast (see Experimental section). The heat capacity measurement for $NpB_2C$ shows an anomaly at 71 K, as an evidence for bulk antiferromagnetism, in good agreement with $T_N$ = 68 K as revealed by magnetic susceptibility studies. There is no anomaly observed for $PuB_2C$ confirming our assignment that $PuB_2C$ is a temperature independent paramagnet.

The insets of Figure 4(a) and 4(b) present $C_p/T$ versus $T^2$ for $NpB_2C$ and $PuB_2C$. Describing the low temperature data according to $C_p/T = \gamma + \beta T^2$ reveals Sommerfeld values $\gamma$ = 14 mJ Pu-mol$^{-1}$ K$^{-2}$ and $\gamma$ = 75 mJ Np-mol$^{-1}$ K$^{-2}$ for $NpB_2C$ and $PuB_2C$, respectively. The relatively large value of $\gamma$ observed for $NpB_2C$ indicates a high density of states, $N(E_F)$, at the Fermi level, $E_F$. The second fitting parameter is the slope $\beta$, which can be used to calculate the Debye temperature $\Theta_D$, i.e., $\beta = 12\pi^4 N k_B / 5\Theta_D^3$, where N is the number of atoms per formula unit, and $k_B$ is Boltzmann's constant. Since $NpB_2C$ and $PuB_2C$ crystallize in the same structure type, and Np and Pu have very similar molar masses, the phonon contribution is expected to be comparable. In reality, the Debye temperature for $NpB_2C$ ($\Theta_D$ = 330 K) is slightly smaller than the Debye temperature estimated for $PuB_2C$ ($\Theta_D$ = 390 K). A possible explanation of this difference is an additional magnetic contribution to the specific heat in the magnetically ordered state that was not taken into consideration for the $NpB_2C$ compound. Correction, if done, should not change much the value of $\gamma$, but might slightly change $\beta$, and consequently the Debye temperature.

The magnetic contribution to the specific heat ($C_{mag}$) for $NpB_2C$ is determined by using the specific heat of the nonmagnetic $PuB_2C$ corrected from its electronic contribution, $\gamma_{PuB2C}$ and is presented in the inset to Figure 5. This magnetic contribution to the specific heat ($C_{mag}$) shows a clear anomaly with the maximum close to $T_N$ = 71 K. The temperature dependence of the magnetic entropy can be then calculated from $S_{mag} = \int \frac{C_{mag}}{T} dT$ and is shown in the main panel of Figure 5. About 90% of the entropy *R* ln2, expected for an S = 1/2 state (a doubly degenerate ground state) is released between 2 K and $T_N$ = 71 K.

## 4. Discussion

Two new members of an actinide based borocarbide family have been synthesized and their magnetic properties, together with the U-based counterpart, are summarized in Table 2. Whereas ß-$UB_2C$ is ferromagnetic, $NpB_2C$ orders antiferromagnetically and $PuB_2C$ shows a temperature independent paramagnetism (TIP). As can be seen from Table 2, replacing an actinide element by a heavier one, causes an increase of the shortest An-An ($d_{An\text{-}An}$) distance. This subtle change might be a clue in understanding magnetic properties in this system. The magnetic behaviour of the three compounds is qualitatively consistent with the ideas of Hill.[24] Compounds for which the inter-atomic spacing between actinide atoms has negligible wave function overlap are anticipated to be magnetic. Considering that Hill limits are 0.34-0.36 nm, 0.325 nm, 0.34 nm for U, Np and Pu, respectively [24], the U, Np and Pu based compounds are in agreement with the Hill criterion. It is worth noting that there are many exceptions to the Hill criterion because the overlap between the actinide and the ligand wavefunctions (not directly related to the interactinide distance) also plays an important role in the hybridization of 5*f* electrons. In the case of plutonium, an additional difficulty is that there are not many examples of magnetically ordered compounds as in uranium or neptunium and the validity of the Hill criterion has been questioned.[25] For example: despite the fact that the Pu-Pu distance for $PuPt_2$ is shorter than 0.34 nm ($d_{Pu\text{-}Pu}$ = 0.331 nm), this compound orders ferromagnetically at $T_C$ = 6 K [26]. It was proposed that details of the 5*f* occupancy is a more determinant criterion for the magnetic ground state of plutonium systems.[25, 27, 28]

The distinctly different electronic transport in both the Np and Pu based compounds is likely the result of significant differences in the electronic structure of both materials: Itinerant Pu may change the number of conduction electrons compared to the more localized Np system and hence can shift the Fermi energy into a region of lower density of states or even into a region of a gap. Hybridisation of conduction electrons with 5*f* electrons, however, can originate a pseudo-gap in the eDOS as a result of many body interactions. This in turn would straightforwardly explain the reduced $\gamma$ value of $PuB_2C$ in combination with the unexpected large electrical resistivity.

## 5. Acknowledgements

We are thankful for Dr Jean Rebizant for his help in this project. The high purity Np and Pu metals required for the fabrication of the title compounds were made available through a loan agreement between Lawrence Livermore National Laboratory and JRC-ITU, in the frame of a collaboration involving LLNL, Los Alamos National Laboratory and the US Department of Energy. We acknowledge the access to infrastructures provided by the European Commission, DG-JRC within its “Actinide User Laboratory” program, and financial support to users provided by the European Commission, DG-Research Contracts No. HPRI-CT-2001-00118, and No. RITA-CT-2006-026176.

## Table 1:

**X-Ray single crystal data[a)] for $NpB_2C$ and $PuB_2C$; $ThB_2C$-type; space group $R\bar{3}m$; No. 166, Z = 9, origin at centre.**

| Parameter/compound | $NpB_2C$ | $PuB_2C$ |
|---|---|---|
| Crystal size<br>*a; c* [nm]<br>$\rho_x$ [$Mgm^{-3}$] | 275×115×50 μm$^3$<br>a = 0.6532(2); c = 1.0769(3)<br>10.162 | 175×70×65 μm$^3$<br>a = 0.6509(2); c = 1.0818(3)<br>10.266 |
| Data collection 295 K, 2Θ range (°)<br>Quadrant<br>Absorption correction<br>Reflections in refinement | 4≤2Θ≤58<br>h=-8/+8; k=-8/+8; l=-14/+14<br>9 Psi-scans (8.42≤Θ≤15.89°)<br>147 ≥3σ(I) of 148 | 4≤2Θ≤54<br>h=0/+8; k=-8/+7; l=-13/+13<br>8 Psi-scans (6.28≤Θ≤13.88°)<br>108 ≥3σ(I) of 127 |
| Number of variables<br>$R_F = \Sigma\|F_0-F_c\|/\Sigma F_0$<br>$R_{Int}$ (total of measured reflections)<br>wR2<br>GOF<br>Extinction (Zachariasen) | 16<br>0.023<br>0.016 (1376)<br>0.039<br>1.38<br>2.32×10$^{-6}$ | 10<br>0.059<br>0.043 (632)<br>0.073<br>2.01<br>4.17×10$^{-7}$ |
| **Np1 (Pu1)** in 3a (0,0,0); occ.<br>$U_{11}=U_{22}=U_{12}$; $U_{33}$; $U_{13}=U_{23}=0.0$ | 1.0<br>0.0061(2); 0.0064(4) | 1.0<br>0.004(1); 0.003(1) |
| **Np2 (Pu2)** in 6c (0,0,z); occ.<br>$U_{11}=U_{22}=U_{12}$; $U_{33}$; $U_{13}=U_{23}=0.0$ | z=0.31225 (7); 1.0<br>0.0051(2); 0.0058(3) | z=0.3124(2); 1.0<br>0.0062(8); 0.0005(9) |
| **B** in 18g (x, 0, ½); occ.<br>$U_{iso}$ | x=0.272 (2); 1.0<br>0.010(3) | x=0.272(8); 1.0<br>0.010(8) |
| **C** in 9d (½,0,½); occ.<br>$U_{iso}$ | 1.00<br>0.014(4) | 1.00<br>0.005(9) |
| Residual density; max; min per Å$^{-3}$ | 1.76; -4.37 | 6.96; -9.00 |
| Interatomic distances [nm]; standard deviations generally < 0.0005 nm | | |
| CN=26 | Np1 – 6 Np2 0.37781<br>2 Np2 0.33626<br>12 B 0.29992<br>6 C 0.26033 | Pu1 – 6 Pu2 0.37648<br>2 Pu2 0.33795<br>12 B 0.29973<br>6 C 0.26041 |
| CN=22 | Np2 – 3 Np2 0.37985<br>3 Np1 0.37781<br>1 Np1 0.33626<br>6 B 0.28691<br>6 B 0.26916<br>3 C 0.24523 | Pu2 – 3 Pu1 0.37852<br>3 Pu2 0.37648<br>1 Pu1 0.33795<br>6 B 0.28668<br>6 B 0.26932<br>3 C 0.24528 |
| CN=9 | B – 2 Np1 0.29992<br>2 Np2 0.28691<br>2 Np2 0.26916<br>2 B 0.17767<br>1 C 0.14893 | B – 2 Pu1 0.29973<br>2 Pu2 0.28668<br>2 Pu2 0.26932<br>2 B 0.17705<br>1 C 0.14841 |
| CN=6 | C – 2 Np2 0.26033<br>2 Np1 0.24523<br>2 B 0.14893 | C – 2 Pu2 0.26041<br>2 Pu1 0.24528<br>2 B 0.14841 |

a) Data collected at RT on an Enraf-Nonius CAD-4; MoKα; 2Θ/ω scans.

Crystal structure data were standardized using Program Typix [29].

**Table 2:**
**Magnetic properties for $AnB_2C$ (An = U, Np, Pu).**

| | **β-$UB_2C$ [a]** | **$NpB_2C$** | **$PuB_2C$** |
|---|---|---|---|
| $\chi_0$ | $600\times10^{-6}$ | $531\times10^{-6}$ | $470\times10^{-6}$ [b] |
| $\Theta_p$ | +75 K | +27 K [c] | - |
| $\mu_{eff}$ | 1.45 $\mu_B$ | 1.65 $\mu_B$ | - |
| Magnetism | $T_C$ = 75 K | $T_N$ = 68 K | TIP |
| The shortest An-An distance (nm) | 0.3352 | 0.3363 | 0.3380 |

[a] ref. 7

[b] possibly due to ferromagnetic impurities

[c] $\chi$(300K)

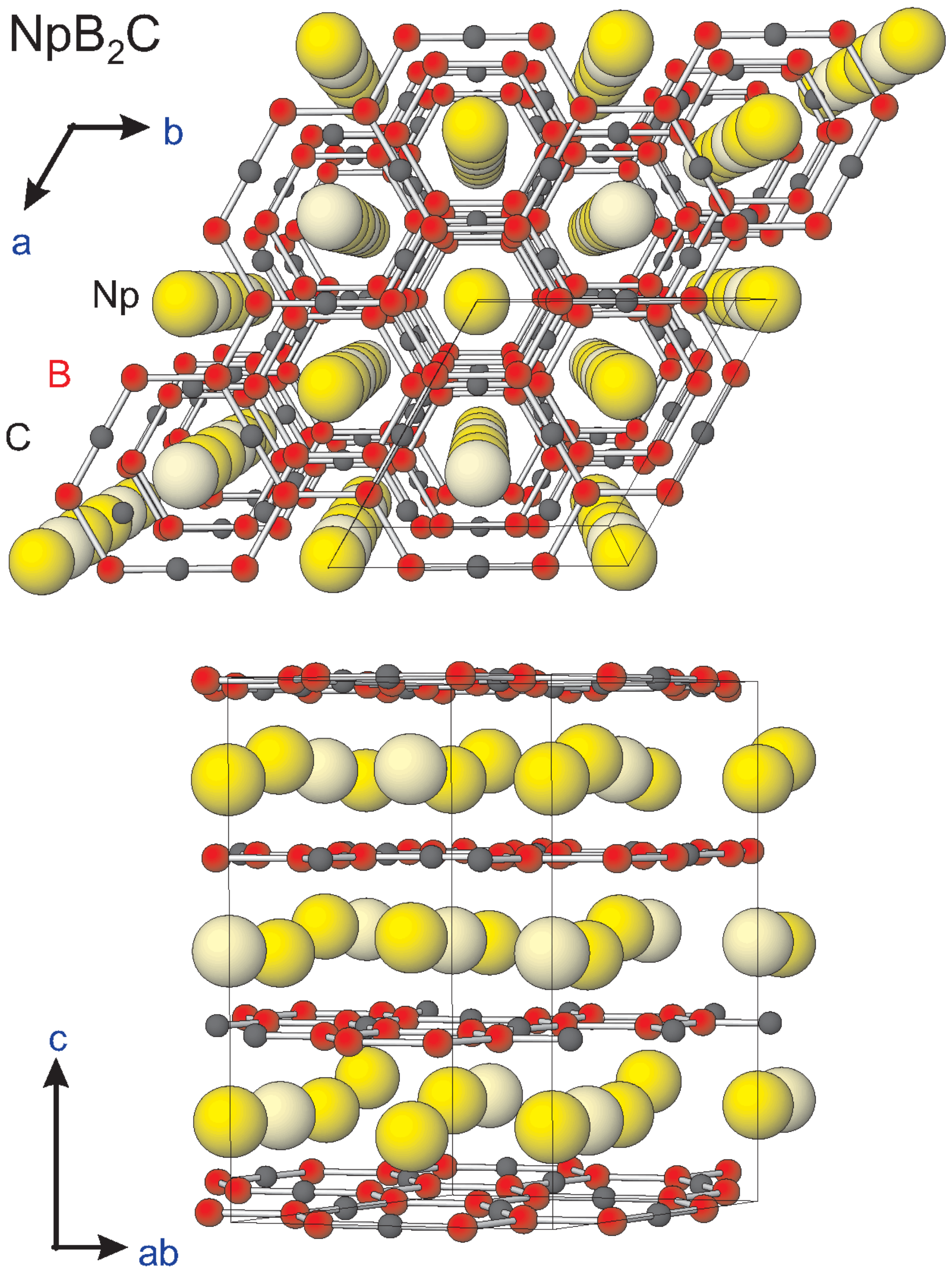

Figure 1: Crystal structure of {Np, Pu}$B_2C$ (ß-$UB_2C$-type) in three-dimensional view. The yellow spheres represent the An1 and An2 atoms, whereas the red and dark spheres represent B and C atoms, respectively. Note the planar non-metal layers that are sandwiched between puckered An-metal layers. The actinide atoms form chains along the *c* axis in the arrangement …-An1-An2-An2-An1-… with the shortest $d_{An\text{-}An}$ distance between An1 and An2 atoms. Note that Tran et al. (ref. 7) proposed U-atoms with moments close to the ab-plane ($\Phi$ = 47° and $\Theta$ = 84°) forming ferromagnetic chains parallel to the hexagonal *c* axis.

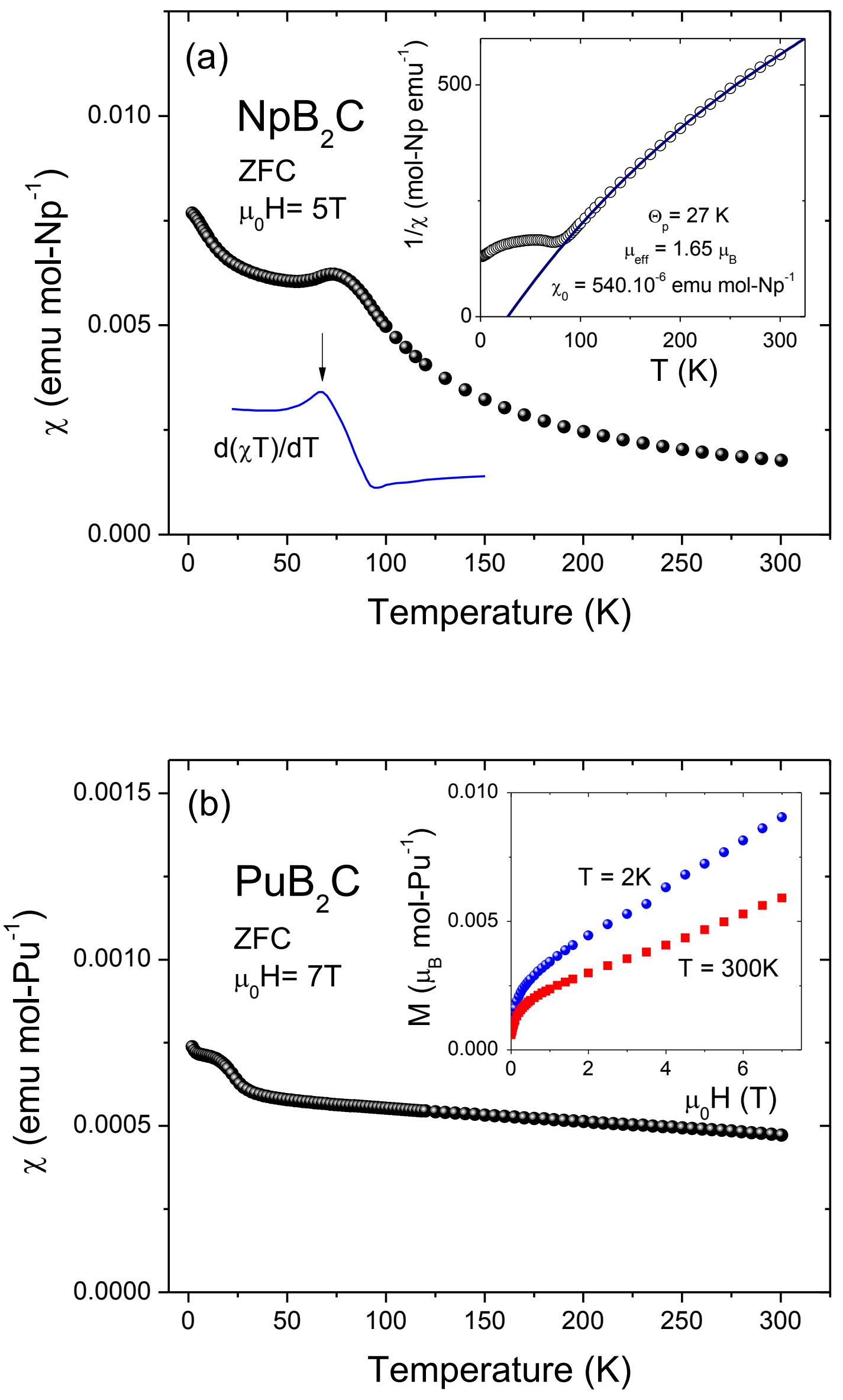

Figure 2: (a) Zero-field-cooled magnetization (solid circles) and d(χT)/dT (solid line) for $NpB_2C$ measured under magnetic field $\mu_0H$ = 5 T. The blue line represents d(χT)/dT in the same temperature scale. The inset shows a Curie-Weiss fit (solid line) to the inverse susceptibility. (b) Zero-field-cooled magnetization for $PuB_2C$ measured under magnetic field $\mu_0H$ = 7 T. The inset shows the field dependence of the magnetization at 2 K (blue circles) and 300 K (red squares).

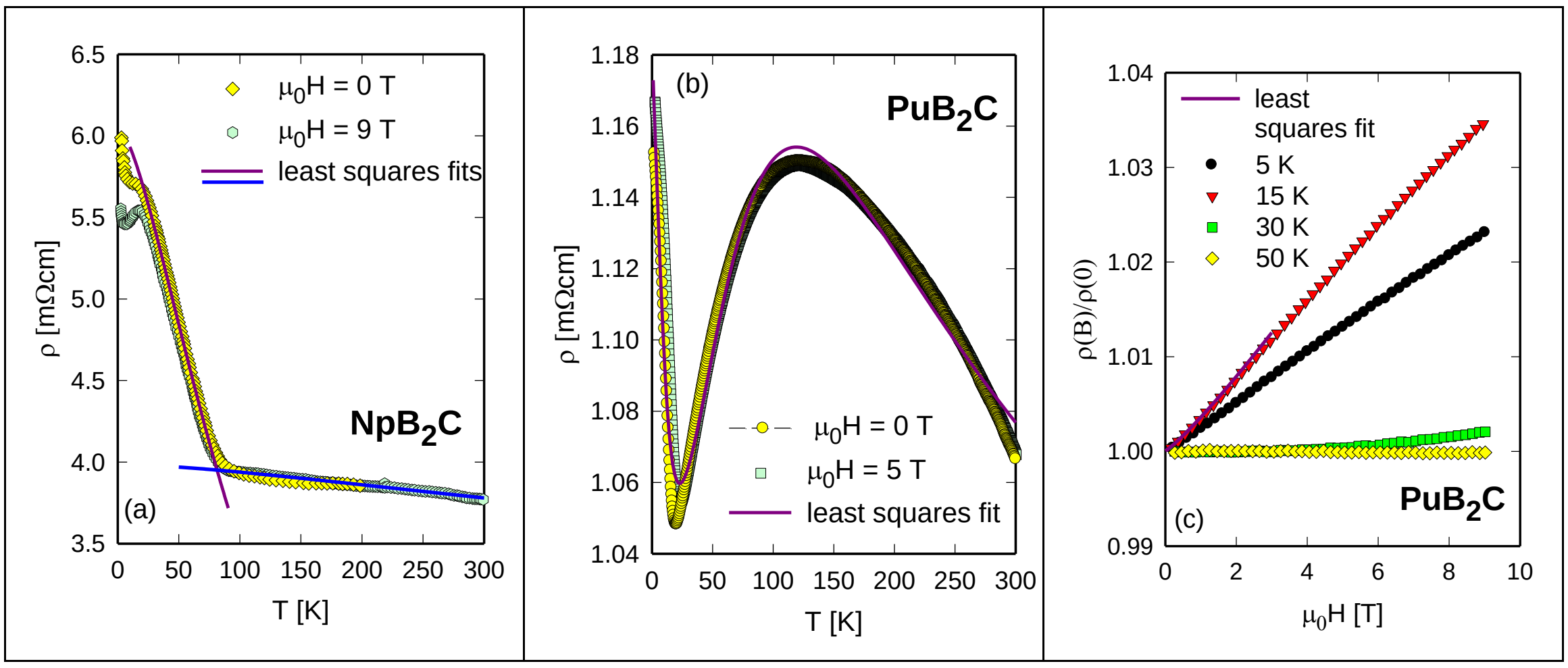

Figure 3: Temperature dependent electrical resistivity of $NpB_2C$ (a) and $PuB_2C$ (b) with and without externally applied magnetic fields. (c) Magnetoresistance for $PuB_2C$ measured at 5 K, 15 K, 30 K and 50 K. The solid lines are least squares fits as explained in the text.

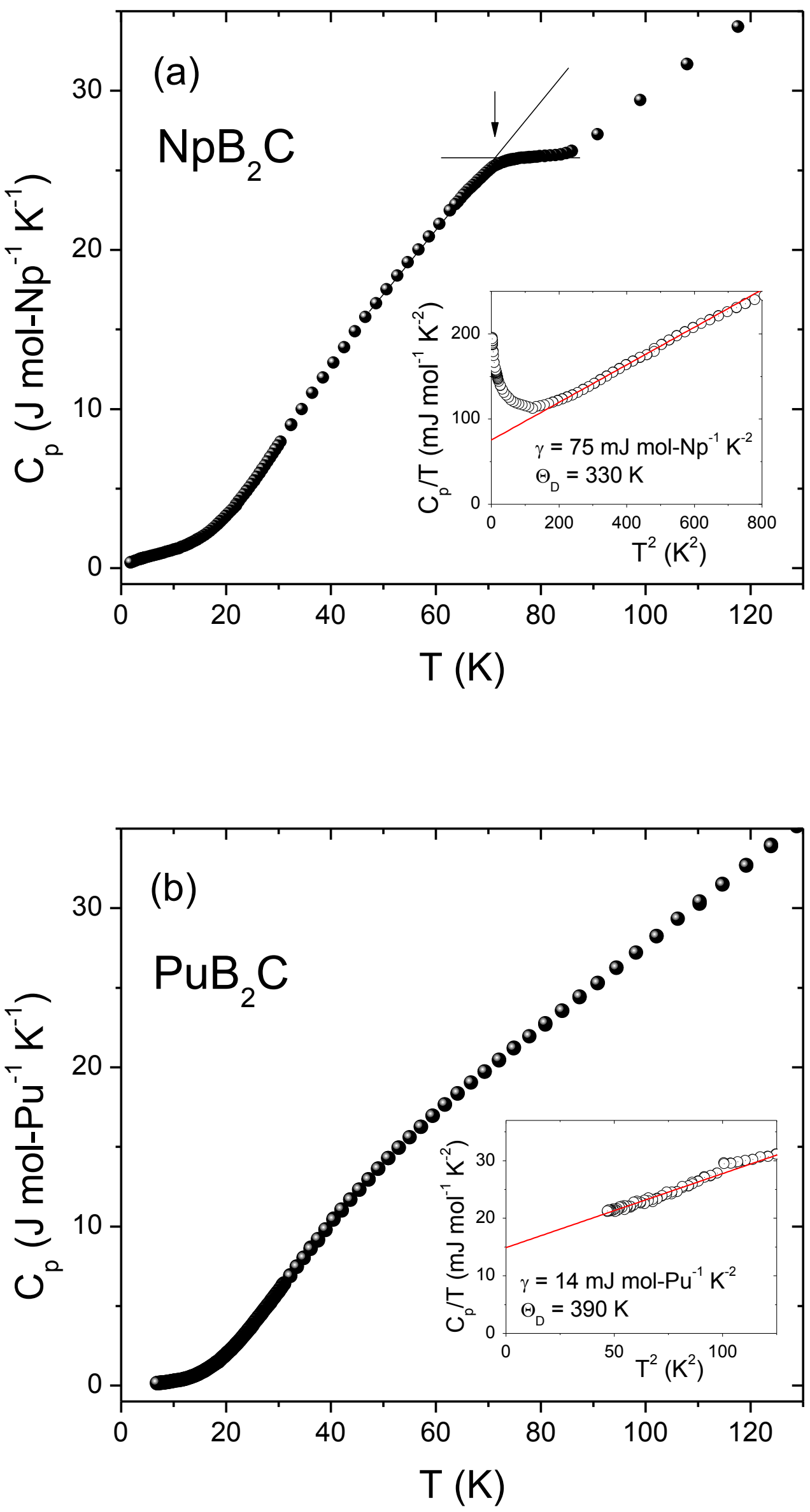

Figure 4: Temperature dependence of the specific heat $C_p$ for $NpB_2C$ (a) and $PuB_2C$ (b) without magnetic field (0 T), presented in the form of $C_p$ vs T (main panels) and $C_p$/T vs $T^2$ (insets). An arrow in part (a) indicates the Neel temperature $T_N$ = 71 K. The red solid lines in the insets represent a fit to the relation $C_p/T = \gamma + \beta T^2$.

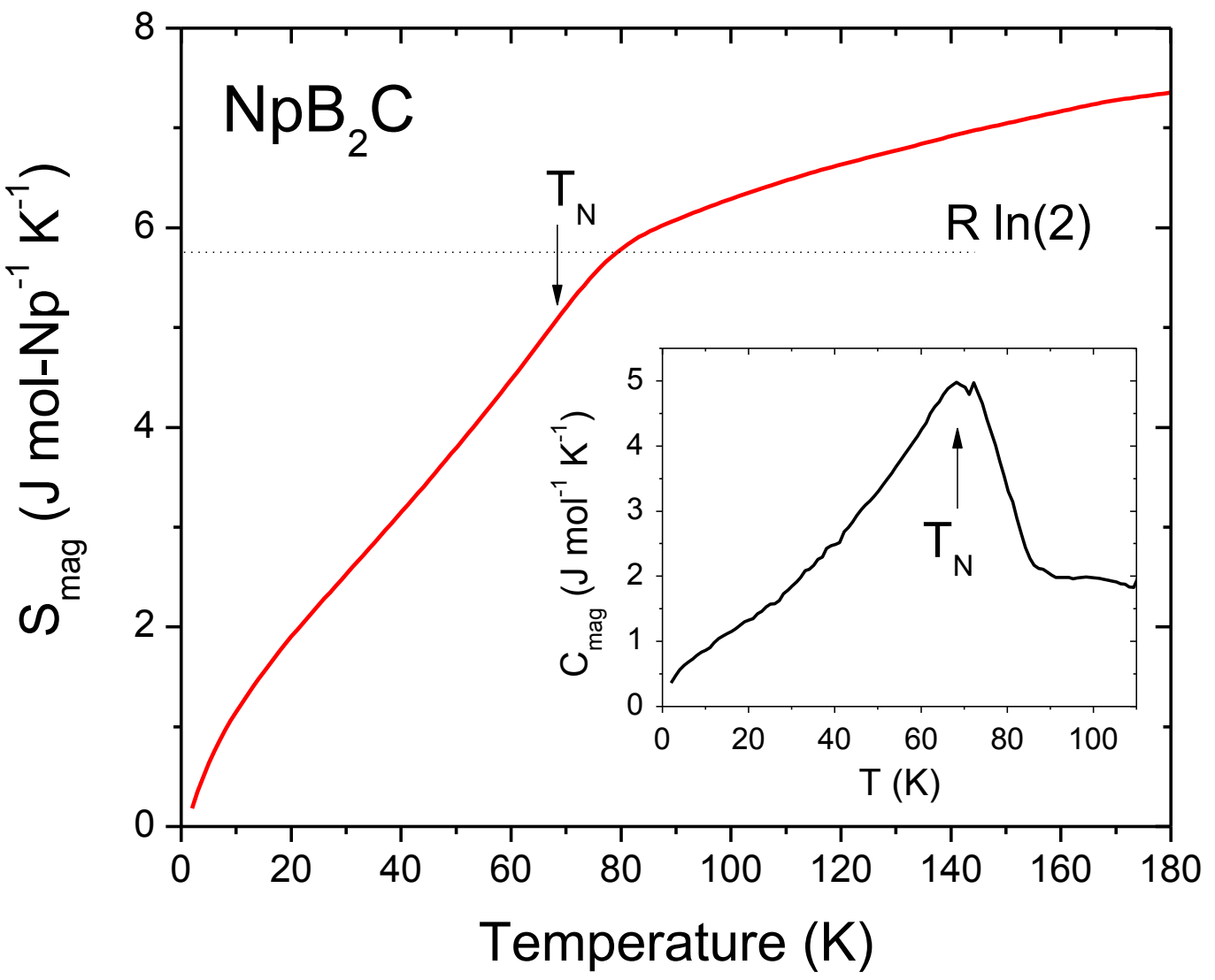

Figure 5: The temperature dependence of the magnetic entropy ($S_{mag}$) for $NpB_2C$. The inset shows magnetic contribution to the specific heat ($C_{mag}$) calculated by the subtraction of the $PuB_2C$ specific heat from the $NpB_2C$ specific heat.